\documentclass[11pt]{article} % use larger type; default would be 10pt

\usepackage{amsmath}
\usepackage{latexsym}
\usepackage{amsfonts}
\usepackage[T2A]{fontenc}

\usepackage[utf8]{inputenc} % set input encoding (not needed with XeLaTeX)

\usepackage{geometry} % to change the page dimensions
\usepackage{graphicx} % support the \includegraphics command and options

\usepackage{booktabs} % for much better looking tables
\usepackage{array} % for better arrays (eg matrices) in maths
\usepackage{paralist} % very flexible & customisable lists (eg. enumerate/itemize, etc.)
\usepackage{verbatim} % adds environment for commenting out blocks of text & for better verbatim
\usepackage{subfig} % make it possible to include more than one captioned figure/table in a single float
\usepackage{fancyhdr} % This should be set AFTER setting up the page geometry
\usepackage{sectsty}
\allsectionsfont{\sffamily\mdseries\upshape} % (See the fntguide.pdf for font help)
\usepackage[nottoc,notlof,notlot]{tocbibind} % Put the bibliography in the ToC
\usepackage[titles,subfigure]{tocloft} % Alter the style of the Table of Contents

\title{The Eikonal Approximation of the Scattering Theory for Fast Charged Particles in a Thin Layer of Crystalline and Amorphous Media  \footnotemark[1] \\
\footnotemark[1]~{submitted to Physics Letters A}}
\author{N.F. Shul'ga \footnotemark[1], V.D. Koriukina \footnotemark[2] \\
\footnotemark[1]~{\textbf{ shulga@kipt.kharkov.ua}} \\ { National Science Center 'Kharkiv Institute of Physics and Techology',} \\{1 Akademichna st., 61108 Kharkiv, Ukraine; } \\{ Karazin Kharkiv National University, 4 Svobody sq., 61102 Kharkiv, Ukraine} \\
\footnotemark[2]~{\textbf{ koriukina@kipt.kharkov.ua}} \\ { National Science Center 'Kharkiv Institute of Physics and Techology',} \\{1 Akademichna st., 61108 Kharkiv, Ukraine }}
\begin{document}
\maketitle

\begin{abstract}
{On the basis of the eikonal approximation of quantum scattering theory, the problem of fast charged particles scattering in a thin crystal when particles fall along one its plane of atoms and in a thin layer of amorphous matter is considered. It is shown that the scattering cross section in this problem, for parameters, which are beyond the scope of application of the Born perturbation theory, differs significantly from the corresponding result of the Born approximation. In this case, the scattering in the transverse to the plane direction is determined mainly by a continuous plane potential, which is widely used in the theory of the channeling phenomenon. The scattering of a particle in the longitudinal direction has features of scattering in a two-dimensional amorphous medium with inhomogeneous density of atoms. The concept of a continuous potential of the crystal plane of atoms in the considered approach appears automatically.
\par}
{KEY WORDS: fast particles scattering in thin media layers, crystal periodical structures, Eikonal and Born approximations of quantum electrodynamics, coherent and incoherent scattering cross section \par}
\end{abstract}

\section{Introduction}

{When fast charged particles move in a crystal near one of its crystallographic axes or planes of atoms, correlations between successive collisions of the particle with the lattice atoms appear. As a result of these correlations, electromagnetic processes in a crystal, such as scattering and radiation, differ significantly from similar processes in an amorphous medium. Consideration of these processes based on the first Born approximation of quantum electrodynamics shows that the cross section for the interaction of particles with crystal lattice atoms in this approximation splits into the sum of the cross sections of coherent and incoherent interaction\cite{MLT72, AIA96}. In this case, in the coherent interaction cross section there are taken into account the correlation characteristics of particle successive collisions with lattice atoms, whereas the incoherent interaction cross section is determined by thermodynamic fluctuations of the atoms positions in the lattice relatively to their equilibrium positions. The cross sections of incoherent interaction in the first Born approximation are slightly different from the corresponding cross sections in an amorphous medium. \par}

{The condition of applicability of the Born approximation in this problem, however, is rapidly destroyed with decreasing angles of particles incidence on a crystal with respect to its crystallographic axes or planes \cite{AIA96, NPK72}. This raises the question how this affects the procedure of separating coherent and incoherent components of the cross section and the kind of these cross section components. This paper is dedicated to this problem analysis on the example of the fast particles scattering when they fall on a thin crystal along one of its crystallographic planes of atoms. The consideration of this problem is carried out on the basis of the eikonal approximation of the quantum scattering theory, which allows one to study electromagnetic processes beyond the scope of application of the Born approximation. It is significant that, on the basis of the eikonal approximation, it is possible to describe the particles scattering both in an amorphous medium and in a crystal using a unified method, since in this approximation there is no specific law of the atoms arrangement in a media.\par}

\section{Scattering in a thin layer of media}

{The process of fast charged particles scattering in a thin layer of matter at small angles can be considered on the basis of the eikonal approximation of quantum electrodynamics. The differential scattering cross section in this approximation is determined by the following formula  \cite{AIA96, NPK72, RJG59,RGN}:

\begin{equation}\frac{d^{2} \sigma }{dq_{\bot}^{2}}=\frac{1}{4 \pi ^{2}} \vert  \int _{}^{}d^{2} \rho  e^{i\frac{\vec{q}_{\perp} \vec{ \rho} }{\hbar}} \left( e^{\frac{i}{\hbar} \Phi \left(  \vec{\rho}  \right) }-1 \right)  \vert ^{2}, \end{equation}

where  \( q_{\bot}= \left( q_{x},q_{y} \right)  \)  is the transverse component of the momentum transmitted by scattered particle to an external field and

\begin{equation}\Phi  \left(  \vec{\rho}  \right) =-\frac{1}{v} \int _{-\infty}^{\infty}dz \, U \left(  \vec{\rho} ,z \right). \end{equation}

 Here, the \textit{z}-axis is parallel to the momenttum of the particle incident on the target,  \(  \vec{\rho} = \left( x,y \right)  \)  are coordinates in the transverse plane,  \( v \)  is particle velocity and  \( U \left(  \vec{\rho} ,z \right)  \) is potential energy of its interaction with atoms of the matter:

 \begin{equation} U \left(  \vec{\rho} ,z \right) = \sum _{n=1}^{N}u \left(  \vec{\rho} - \vec{\rho} _{n},z-z_{n} \right), \end{equation}

where  \( u \left(   \vec{\rho} -  \vec{\rho} _{n},z-z_{n} \right)  \)  is potential energy of particle interaction with a separate atom of the matter,  \(\vec{ r}_{n}= \left(   \vec{\rho} _{n},z_{n} \right)  \)  is atom position in the media and  \( N \)  is amount of atoms in the media. \par}

 {For particle scattering in a thin layer of amorphous media, formula (1) should be averaged over equally probable positions of atoms in the medium. Moreover, in the Born approximation, when the scattering phase  \( \lvert  \Phi  \left(  \vec{\rho}  \right)/ \hbar \rvert  \)  is small compared to unity, the mean value of the scattering cross section (1) is the sum of the cross sections of a particle scattering on all atoms of the medium :

 \begin{equation} \frac{d^{2} \sigma ^{ \left( N \right) }}{dq_{\bot}^{2}}=N\frac{d^{2} \sigma ^{ \left( 1 \right) }}{dq_{\bot}^{2}}\end{equation}

With increasing target thickness, the scattering phase  \(  \Phi  \left(  \vec{\rho}  \right)  \)  increases. Moreover, if  \(  \vert  \Phi  \left(  \vec{\rho}  \right)  \vert  \gg \hbar \) , then as a result of averaging we obtain the following expression for the scattering cross section (1) in a thin layer of amorphous media \cite{AIA96}:

 \begin{equation}  \langle \frac{d^{2} \sigma }{dq_{\perp}^{2}} \rangle =S\frac{1}{ \pi \langle q_{\perp}^2 \rangle}  exp \left( -\frac{\vec{q}_{\perp}^{2}}{\langle q_{\perp}^2 \rangle}  \right) \end{equation}

 where  \( S \)  is target cross section and  \( \langle q_{\perp}^2 \rangle \)  is average squared value of transmitted momentum:

 \begin{equation} \langle q_{\perp}^2 \rangle =Ln \int _{}^{}d \sigma _{1} \left( q_{\perp} \right) q_{\perp}^{2}\end{equation}

 Deriving (5), we used the fact that  \( N=nLS \) , where  \( n \)  is the density of atoms in the target and  \( L \)  is the target thickness. \par}

{ In a crystal, atoms form a regular structure, so that the particles scattering in a crystal can differ significantly from scattering in an amorphous medium. The positions of atoms in the crystal lattice have some small deviation  \(\vec{ u}_{n} \)  from their equilibrium positions  \(\vec{ r}_{n}^{0} \) :

\begin{equation}\vec{ r}_{n}=\vec{r}_{n}^{0}+\vec{u}_{n} \end{equation}

 This deviation is due to thermal vibrations of atoms in the crystal. The particles scattering in such structure depends on the orientation of the crystallographic axes relatively to the direction of the incident beam.
\par}

\section{Scattering in the Field of Separate Crystalline Plane of Atoms}

 {Let us consider the fast particles scattering in a crystal studying one of the simplest variants of the crystallographic axes orientation relatively to the incident beam, when the particles fall on the crystal along the parallel crystal planes of atoms. We analyze the problem on the basis of the eikonal approximation of the quantum scattering theory, which allows one to exceed the scope of application of the Born perturbation theory. The usage of the eikonal approximation in this problem is possible if the particle motion in a crystal is close to a straight-line motion. This requires small crystal thickness along the particle movement direction compared with the length that particle passes during the period of its oscillation in the interplanar potential. Under these conditions, the corrections of the eikonal scattering phase  \(  \Phi  \left(  \vec{\rho}  \right)  \)  are small \cite{AIA96}.\par}

{The features of the particle (plane wave) scattering in a crystal in the given problem are determined both with the features of its interaction with atoms of separate crystal plane, and with features of interference effects of scattering on different crystal planes of atoms. In this paper, we focus on the study of the features of particle scattering in the field of a separate crystal plane of atoms (see Fig. 1), paying special attention to the issues of theoretical considerations beyond the scope of application of the Born perturbation theory. There is studied the simplest version of this problem, in assumption that the atoms positions in crystalline plane are equiprobable with density  \( n_{\perp}=N_{p}/S \) , where  \( N_{p} \)  is the number of atoms in the plane and  \( S \)  is its area. In this case, the equiprobability of atoms positions in the plane is equivalent to a two-dimensional amorphous medium.\par}

%%%%%%%%%%%%%%%%%%%% Figure/Image No: 1 starts here %%%%%%%%%%%%%%%%%%%%

\begin{figure}
	\centering
		\includegraphics[scale=.6]{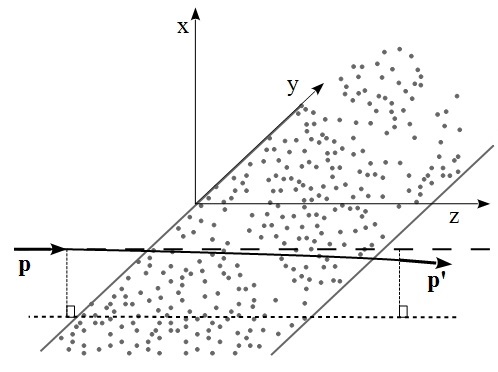}
	\caption{ Scattering in the field of a separate crystal plane of atoms}
	\label{FIG:1}
\end{figure}

%%%%%%%%%%%%%%%%%%%% Figure/Image No: 1 Ends here %%%%%%%%%%%%%%%%%%%%

 Inserting (7) into (2), we obtain the following expression for the scattering phase in cross section (2):

\begin{equation}\Phi  \left(  \vec{\rho}  \right) = \sum _{n=1}^{N_{p}} \chi  \left(  \vec{\rho} - \vec{\rho} _{n} \right)\end{equation}

{where   \(  \vec{\rho} _{n}= \left( u_{x,n},y_{n} \right)  \)  is atom position in  \(  \left( x,y \right)  \) -plane,  \( x \) -axis is orthogonal to   \(  \left( y,z \right)  \) -plane of atoms in crystal, along this plane the particle moves (see Fig.\ref{FIG:1}) and

\begin{equation}\chi  \left(  \vec{\rho} - \vec{\rho} _{n} \right) =-\frac{1}{v} \int _{-\infty}^{\infty}dz \, u \left(  \vec{\rho} - \vec{\rho} _{n},z-z_{n} \right)\end{equation}

 Here  \( z \) -axis is parallel to the incident particle momentum and  \( u \left(  \vec{\rho} - \vec{\rho} _{n},z-z_{n} \right)  \)  is potential energy of particle interaction with a separate atom of a plane located at the point with  \( \vec{r}_{n}= \left(  \vec{\rho} _{n},z_{n} \right)  \)  coordinates. Let us notice that  \(  \chi  \left(  \vec{\rho} - \vec{\rho} _{n} \right)  \)  value does not depend on  \( z_{n} \) . \par}

 {Inserting (8) into (1), we find out that differential cross section of scattering at nonzero angles (1) becomes the following:

\begin{equation}\frac{d^{2} \sigma }{dq_{\perp}^{2}}=\frac{1}{4 \pi ^{2}} \int _{}^{}d^{2} \rho  d^{2} \rho ' e^{i\frac{\vec{q}_{\perp} \left(  \vec{\rho} - \vec{\rho} ' \right) }{\hbar}}e^{\frac{i}{\hbar}  \left[ \Phi \left(  \vec{\rho}  \right) - \Phi \left(  \vec{\rho} ' \right)  \right] }\end{equation}
\par}

{The scattering cross section (10) should be averaged over the each atom position in the plane. Assuming for simplicity that the atoms distribution in the plane along the \textit{y}-axis is equiprobable with the distribution function
\begin{equation} g \left( y_{0} \right) =\frac{1}{L_{y}}, \qquad 0 \leq y_{0} \leq L_{y}\end{equation}

 and that the position deviation of each atom  \( u_{x, n} \)  along the  \( x \) -axis relative to its equilibrium position in the crystal plane is Gaussian:

\begin{equation} f \left( u_{x,n} \right) =\frac{1}{\sqrt[]{2 \pi\langle u^2 \rangle }}e^{-\frac{u_{x,n}^{2}}{2 \langle u^2 \rangle }}\end{equation}

 we obtain the following expression for the average value of the function in (10):

\begin{eqnarray} &  \langle e^{\frac{i}{\hbar}  \left[ \Phi \left(  \vec{\rho}  \right) - \Phi \left(  \vec{\rho} ' \right)  \right] } \rangle =  \prod_{n=1}^{N_{p}} \int _{}^{}du_{x,n}f \left( u_{x,n} \right)  \int _{0}^{L_{y}}dy_{0,n}g \left( y_{0,n} \right) e^{\frac{i}{\hbar}  \left[  \chi  \left(  \vec{\rho} - \vec{\rho} _{n} \right) - \chi  \left(  \vec{\rho} '- \vec{\rho} _{n} \right)  \right] }\end{eqnarray}
 where  \( L_{y} \)  is the size of the crystal plane along the \textit{y}-axis and  \(\bar{ u^2} \)  is the atom average squared displacement along the \textit{x}-axis relatively to the equilibrium position.
\par}

{All \textit{n}-th multipliers in (13) are the same, therefore

\begin{eqnarray} &  \langle e^{\frac{i}{\hbar}  \left[ \Phi \left(  \vec{\rho}  \right) - \Phi \left(  \vec{\rho} ' \right)  \right] }  =  \left[  \int _{}^{}du_{x}f \left( u_{x} \right)   \int _{0}^{L_{y}}dy_{0}g \left( y_{0} \right) e^{\frac{i}{\hbar}  \left[  \chi  \left(  \vec{\rho} - \vec{\rho} _{0} \right) - \chi  \left(  \vec{\rho} '- \vec{\rho} _{0} \right)  \right] } \right] ^{N_{p}} & \end{eqnarray}
 where  \(  \vec{\rho} _{0}= \left( u_{x},y_{0} \right)  \)
\par}

{ For further speculations, we write the relation (14) in the following form:
\begin{eqnarray} & \langle e^{\frac{i}{\hbar}  \left[ \Phi \left(  \vec{\rho}  \right) - \Phi \left(  \vec{\rho} ' \right)  \right] } \rangle =exp \left\{ N_{p} ln \left[  \int _{}^{}du_{x}f \left( u_{x} \right)   \int _{0}^{L_{y}}dy_{0}g \left( y_{0} \right) e^{\frac{i}{\hbar}  \left[  \chi  \left(  \vec{\rho} - \vec{\rho} _{0} \right) - \chi  \left(  \vec{\rho} '- \vec{\rho} _{0} \right)  \right] } \right]  \right\} & \end{eqnarray}
\par}

{ For large values of \(N_p \), the main contribution to the scattering cross section (10) will be made by the values of  \(  \vec{\rho}  \) , which are close to  \(  \vec{\rho} ' \) . In this case, in (15) the expansion with  \(  \left(  \vec{\rho} - \vec{\rho} ' \right)  \)  small parameter can be made. We carry out this expansion in two stages, firstly making the expansion with  \(  \left[  \chi  \left(  \vec{\rho} - \vec{\rho} _0 \right) - \chi  \left(  \vec{\rho} '- \vec{\rho} _0 \right)  \right]/\hbar \)  small value. Preserving the quadratic terms in the expansion with this parameter, we find out that

\begin{eqnarray} & \langle e^{\frac{i}{\hbar}  \left[ \Phi \left(  \vec{\rho}  \right) - \Phi \left(  \vec{\rho} ' \right)  \right] } \rangle  \approx exp \left\{ N \left[ \frac{i}{\hbar} \langle   \chi - \chi'  \rangle -\frac{1}{2\hbar^{2}} \langle \left(  \chi - \chi ' \right) ^2 \rangle+\frac{1}{2\hbar^{2}}  \langle  \chi - \chi ' \rangle  ^2+ \ldots  \right]  \right\}\end{eqnarray}

where the pair of angle brackets means the averaging procedure of the corresponding functions over  \( u_{x} \)  and  \( y_{0} \) ,

\begin{eqnarray} & \langle \chi - \chi ' \rangle= \int _{}^{}du_{x}f \left( u_{x} \right)  \int _{0}^{L_{y}}\frac{dy_{0}}{L_{y}} \left[  \chi  \left(  \vec{\rho} - \vec{\rho} _{0} \right) - \chi  \left(  \vec{\rho} '- \vec{\rho} _{0} \right)  \right] \end{eqnarray}
\par}

{The number of atoms  \(  N_p  \) in the crystal plane  \(  \left( y,z \right)  \) , along which the particles fall, is proportional to the density of atoms in this plane  \( n_{\perp} \)  and its surface area  \( S=L_{y}L_{z} \) :
\begin{equation} N_{p}=n_{\perp}L_{y}L_{z}\end{equation}
 where  \( L_{y} \)  and  \( L_{z} \)  are the plane sizes along the \textit{y} and \textit{z} axes, therefore, as  \( L_{y} \rightarrow \infty \) , the last term in exponent (16) can be neglected.
\par}

{Now we make expansion with  \(\vec{ \Delta} = \vec{\rho} - \vec{\rho} ' \)  small value in  \(  \langle  \chi - \chi' \rangle  \)  and  \(  \langle \left(  \chi - \chi' \right)  ^2 \rangle \)  in (16). It is easy to verify that, taking into account the $ \vec{ \Delta} $  quadratic terms in this expansion leads to the following relations:

\begin{eqnarray}& \langle \chi - \chi' \rangle \approx \frac{1}{L_{y}} \int _{-\infty}^{\infty}du_{x}f \left( u_{x} \right)  \int _{0}^{L_{y}}dy_{0} \left[  \Delta _{x}\frac{ \partial }{ \partial x^{'}}+\frac{ \Delta _{x}^{2}}{2}\frac{ \partial ^{2}}{ \partial x'^{2}} \right]  \chi  \left( x^{'}-u_{x},y_{0} \right); &  \\
&\langle \left(  \chi - \chi' \right)  ^2 \rangle \approx \frac{1}{L_{y}} \int _{-\infty}^{\infty}du_{x}f \left( u_{x} \right)   \int _{0}^{L_{y}}dy_{0} \left[  \Delta _{x}^{2} \left( \frac{ \partial  \chi  \left( x^{'}-u_{x},y_{0} \right) }{ \partial x^{'}} \right) ^{2}+ \Delta _{y}^{2} \left( \frac{ \partial  \chi  \left( x^{'}-u_{x},y_{0} \right) }{ \partial y_{0}} \right) ^{2} \right] & {}  \nonumber \end{eqnarray}

We used here the fact that the  \(  \Delta _{y} \)  linear terms containing  \(  \partial  \chi  \left( x^{'}-u_{x},y_{0} \right) / \partial y_{0} \)  disappear as a result of integration over  \( y_{0} \) . The similar situation occurs in the term  \( \langle \chi - \chi' \rangle \)  containing the second derivatives with  \( y_{0} \) .
\par}

{ Inserting (16) into (10) and taking into account (19), we obtain the following expression for the scattering cross section:

\begin{eqnarray}& \langle \frac{d^{2} \sigma }{dq_{\perp}^{2}} \rangle =\frac{1}{4 \pi ^{2}} \int _{}^{}dx^{'}dy^{'}  \int _{}^{}d \Delta _{x}d \Delta _{y} e^{i\frac{q_{\perp} \Delta }{\hbar}}exp \left\{ i\frac{ \Delta _{x}}{\hbar}A+\frac{ \Delta _{x}^{2}}{2\hbar^{2}} \left( iB-D \right) -\frac{ \Delta _{y}^{2}}{2\hbar^{2}}F \right\}&\end{eqnarray}

where

\begin{eqnarray}
& A=n_{\perp}L_{z} \int _{-\infty}^{\infty}du_{x}f \left( u_{x} \right)  \int _{0}^{L_{y}}dy_{0}\frac{ \partial  \chi  \left( x^{'}-u_{x},y_{0} \right) }{ \partial x^{'}}; & {} \nonumber\\
 & B=n_{\perp}L_{z} \int _{-\infty}^{\infty}du_{x}f \left( u_{x} \right)  \int _{0}^{L_{y}}dy_{0}\frac{ \partial ^{2}  \chi  \left( x^{'}-u_{x},y_{0} \right)}{ \partial x'^{2}}; & \\
& D=n_{\perp}L_{z} \int _{-\infty}^{\infty}du_{x}f \left( u_{x} \right)  \int _{0}^{L_{y}}dy_{0} \left( \frac{ \partial  \chi  \left( x^{'}-u_{x},y_{0} \right) }{ \partial x^{'}}   \right) ^{2}; & {} \nonumber\\
& F=n_{\perp}L_{z} \int _{-\infty}^{\infty}du_{x}f \left( u_{x} \right)  \int _{0}^{L_{y}}dy_{0} \left( \frac{ \partial   \chi  \left( x^{'}-u_{x},y_{0} \right) }{ \partial y^{'}_0}  \right) ^{2}. & {}   \nonumber
\end{eqnarray}
\par}

 {After integration over  \(  \Delta _{x} \)  and  \(  \Delta _{y} \)  we find out that
\begin{eqnarray} \langle \frac{d^{2} \sigma }{dq_{\perp}^{2}} \rangle =\frac{L_{y}}{4 \pi ^{2}} \int _{}^{}dx^{'}\sqrt[]{\frac{2 \pi }{iB-D}}e^{- \frac{ \left( q_{x}+A \right) ^{2}}{2 \left( iB-D \right) }}\sqrt[]{\frac{2 \pi }{F}}e^{- \frac{q_{y}^{2}}{F}} \end{eqnarray}
\par}

{ We used here the fact that the integrand (20) does not depend on  \( y^{'} \) . Integration over  \( y^{'} \)  gives  \( L_{y} \) . To calculate the integral over  \( x^{'} \) , we use the stationary phase method, according to which the main contribution to this integral is made by the  \( x^{'} \)  values, which are close to the  \( \tilde{ x} \)  value determined with relation:
\begin{equation}q_{x}=-A \left( \tilde{x} \right)\end{equation}
\par}

{Considering mentioned above, for the convergence of the integral over  \( x^{'} \)  in (22), the function  \( A=A \left( x^{'} \right)  \)  should be expanded in series with  \(  \left( x^{'}-\tilde{x} \right)  \)  upto linear terms, while other functions in the integrand should be taken with  \( x^{'}=\tilde{x} \). After integration over  \(  \left( x^{'}-\tilde{x} \right)  \) , we obtain the following expression for the scattering cross section:
\begin{eqnarray}\langle \frac{d^{2} \sigma }{dq_{\perp}^{2}} \rangle =\frac{L_{y}}{2 \pi }\frac{1}{ \vert \frac{dq_{x} \left( \tilde{x} \right) }{d \tilde{x}} \vert }e^{- \frac{q_{y}^{2}}{ \langle q_{y}^{2}  \rangle }}\sqrt[]{\frac{ \pi }{ \langle q_{y}^{2} \rangle }}\end{eqnarray}

 where  \( q_{x} \left( \tilde{x} \right) =-A \left( \tilde{x} \right)  \)  and  \(  \langle q_{y}^{2} \rangle =2F \left( \tilde{x} \right)  \) 
\par}

\section { Discussion}

{The obtained results show that the atoms ordering in the target leads to significant changes in the features of the particles scattering in the substance. On the basis of the eikonal approximation, it is possible to consider from a single point of view the particles scattering in both amorphous and ordered media.\par}

{ When a particle is scattered in the field of a separate atomic crystal plane, the  \(  \langle  \chi - \chi ^{'} \rangle  \)  value in (16) is determined by the continuous potential of the crystal plane of atoms, which is widely used to describe the fast particles motion in planar channeling case \cite{AIA96, JLi,DSG}. This potential lets us to take into account the correlation properties of particle scattering during its successive collisions with atoms of the plane. The continuous potential is determined in (17), taking into account the thermal vibrations of atoms along the \textit{x}-axis. Thus, in the considered approach, the concept of a continuous potential of the crystal plane of atoms appears naturally.\par}

{ The quadratic combinations of the  \(  \chi  \)  and  \(  \chi ^{'} \)  values in (16) determine the scattering incoherent effects associated with the influence of the potential fluctuations of the crystal plane relatively to its average value on the scattering. In an amorphous medium (in this case  \(  \langle  \chi - \chi ^{'} \rangle =0 \) ),\ the inclusion of these quadratic terms leads to effects associated with multiple scattering on substance atoms, and the scattering is described by the Gaussian distribution of  deflection angles of scattered particles (5). We note that if we set  \( \langle u^{2} \rangle \rightarrow \infty \) (an amorphous media case), the values of \textit{A} and \textit{B} in (22) turn into zero. In this case, formula (22) becomes equal to the corresponding result of the theory of particles multiple scattering in matter (5). Thus, using the eikonal approximation, it is possible to consider both the correlation effects in scattering and the effects associated with particles multiple scattering in a substance from a unified point of view.\par}

{ Formula (24) shows that the scattering cross section of a particle in the field of a thin separate crystal plane of atoms is a product of two multipliers determining the scattering in the field of the corresponding continuous potential of the crystal plane of atoms along the \textit{x}-axis and one-dimensional multiple scattering along the \textit{y}-axis. The first of these multipliers coincides with the corresponding result of the classical theory of scattering in the field of a continuous potential of a thin crystal plane (the reduced Planck constant $\hbar$  in this multiplier vanishes as a result of using the stationary phase method to calculate the integral over \textit{x'} in (22)). The second multiplier is the Gaussian distribution of scattering angles along the \textit{y}-axis in a medium with a non-uniform atoms density distribution. The mean value of the squared scattering angle along the \textit{y}-axis in this case is determined by the point of the stationary phase  \( \tilde{x} \) , which, in its turn, is determined by the scattering angle value along the \textit{x}-axis. Thus, the particles scattering characteristics along the \textit{y} and \textit{x} axes in this case turn out to be interconnected despite the different mechanisms of particle scattering along the \textit{y} and \textit{x} axes.\par}

{ We note that the formula for the scattering cross section (24) differs from the corresponding result of the Born approximation. In the Born approximation, the exponent in (16) must be expanded in series in powers of  \(  \chi  \) . Taking into account the quadratic terms of the expansion in  \(  \chi  \) , we obtain the following expression for the scattering cross section (10):

\begin{eqnarray}& \langle \frac{d^{2} \sigma }{dq_{\perp}^{2}} \rangle  \approx \frac{1}{4 \pi ^{2}} \left\{ N \left( 1-e^{- \frac{q_{\perp}^{2} \langle u^2 \rangle }{2}} \right)  \vert  \chi _{q_{\perp}} \vert ^{2} +N^{2}e^{- \frac{q_{\perp}^{2}\langle u^2 \rangle }{2}} \vert  \chi _{q_{\perp}} \vert ^{2}  \right\}, &   \\
&\chi _{q_{\perp}}= \int _{}^{}d^{2} \rho e^{i\frac{\vec{q} \vec{\rho} }{\hbar}} \chi  \left(  \vec{\rho}  \right) 
 &   \nonumber \end{eqnarray}
\par}

{The term in (25), proportional to \textit{N}, is a cross section of incoherent scattering. The term proportional to  \( N^{2} \)  is a coherent scattering cross section.\par}

{ Formula (25) has the same structure as the corresponding result of the Ter-Mikaelyan theory \cite{MLT72, TM53} for ultrarelativistic electrons coherent radiation in a crystal, obtained on the basis of the Born perturbation theory.\par}

{ The difference between formulas (24) and (25) is due to the following reasons. In deriving (24), it was assumed that the main contribution to the scattering cross section is made by the values of  \(  \vec{\rho}  \) , which are close to  \(  \vec{\rho} ' \) . The conditions of applicability of the Born approximation in this problem are opposite to the conditions under which the derivation of formula (24) was carried out.\par}

{ The suggested method for obtaining the scattering cross section of fast charged particles in a crystal, based on the eikonal approximation of the quantum scattering theory, can be generalized to the case of particles scattering in the field of periodically located crystal planes of atoms, as well as to the cases of other orientations of the crystallographic axes relatively to  direction of the incident beam motion. These cases, however, require special consideration due to more complex arrangement of atoms in the target.\par}

{The main ideas of this paper can also be generalized to the case of thicker crystals, in which it is necessary to apply the theory beyond the eikonal approximation. Of particular interest in this regard are the possibilities of applying geometric optics methods to this problem \cite{NFS19}. This aspect, however, also requires special consideration. 
\par}

\section*{Acknowledgements}
{The work was partially supported by the projects L10/56-2019 and C-2/50-2018 of the National Academy of Science of Ukraine.
\par}

\end{document}